\let\SchwingerKernelLabel\label
\documentclass[
  aps,  prd,reprint,superscriptaddress,
nofootinbib,floatfix,amsmath,
  amssymb,longbibliography]{revtex4-2}
\makeatletter
\@ifl@t@r\fmtversion{2023-06-01}{%
  \let\label\SchwingerKernelLabel
  \let\ltx@label\label
}{}
\makeatother
\usepackage[colorlinks, linkcolor=darkblue, citecolor=darkblue, urlcolor=darkblue, linktocpage]{hyperref}

\usepackage{amsfonts,stmaryrd}
\usepackage{graphicx}
\usepackage[export]{adjustbox}
\usepackage{enumerate}
\usepackage{cancel}
\usepackage{color}
\usepackage{tikz}
\usepackage[caption=false,font=normalsize]{subfig}
\usepackage{etoolbox}
\makeatletter
\patchcmd{\@makecaption}{\small\rmfamily}{\normalsize\rmfamily}{}{
  \PackageError{schwinger}{Could not set the REVTeX caption font}{}
}
\makeatother
\usepackage{tikz-cd}
\usetikzlibrary{arrows,backgrounds}
\usepackage{url}
\definecolor{darkblue}{rgb}{0.1,0.1,.7}
\usepackage{textcomp}
\usepackage{amsthm}
\usepackage{bbold}
\usepackage{mathtools}
\usepackage{multirow}
\usepackage{dsfont}
\usepackage{braket}
\usepackage{leftindex}
\usepackage[normalem]{ulem}
\usepackage{stackrel}
\usepackage[vcentermath]{youngtab}

\usepackage{amstext}
\usepackage{array}
\newcolumntype{L}{>{$}l<{$}}
\newcolumntype{C}{>{$}c<{$}}

\def\p{\partial}

\newcommand{\eps}{\epsilon}

 \newcommand{\E}{|\langle\mathcal E\rangle|}

\newcommand{\calL}{\mathcal{L}}

\def\be{\begin{equation}}
\def\ee{\end{equation}}

\definecolor{redcurve}{rgb}{1.0,0.3907,0.3056}
\definecolor{bluecurve}{rgb}{0.0006,0.6336,1.0}
\definecolor{greendot}{rgb}{0.113,0.6938,0.0017}

\usepackage{microtype}
\makeatletter
\def\SchwingerShowHyphensCompat#1#2{%
  \def\MT@temp##1##2{%
    \MT@ifdefined@n@TF{showhyphens }{%
      \MT@exp@cs\patchcmd{showhyphens }
        {\color@begingroup}{\color@begingroup\pdfadjustspacing\z@}
        {}{\PackageError{schwinger}{Could not patch showhyphens}{}}%
    }{%
      \patchcmd{\showhyphens}
        {\color@begingroup}{\color@begingroup\pdfadjustspacing\z@}
        {}{\PackageError{schwinger}{Could not patch showhyphens}{}}%
    }}}
\newif\ifSchwingerLegacyMicrotype
\@ifpackagelater{microtype}{2025/07/08}{}{\@ifl@t@r\fmtversion{2025-06-01}{\SchwingerLegacyMicrotypetrue}{}}
\ifSchwingerLegacyMicrotype
  \microtypesetup{nopatch=footnote}
  \patchcmd{\MT@setup@expansion}
    {\MT@ifdefined@n@TF{showhyphens }}{\SchwingerShowHyphensCompat}
    {}{\PackageError{schwinger}{Could not apply the microtype compatibility fix}{}}
\fi
\makeatother
\usepackage{booktabs}
\makeatletter
\patchcmd{\do@startpage}
  {\check@deferlist@stuck\do@startpage}
  {\fcolmade@sw
    {\global\let\@deferlist@postshipout\@empty}
    {\check@deferlist@stuck\do@startpage}}
  {}{\PackageError{schwinger}{Could not repair the REVTeX float-page check}{}}
\makeatother

\newcommand{\AffiliationSymbol}[1]{%
  \ifcase#1\relax
  \or\ensuremath{\tau}%
  \or\ensuremath{m}%
  \else#1%
  \fi
}
\makeatletter
\patchcmd{\@affilcomma}
  {#1\@ifx}
  {\AffiliationSymbol{#1}\@ifx}
  {}{\PackageError{schwinger}{Could not set the author affiliation markers}{}}
\patchcmd{\@affil@script}
  {\def\@thefnmark{#1}}
  {\def\@thefnmark{\AffiliationSymbol{#1}}}
  {}{\PackageError{schwinger}{Could not set the institution affiliation markers}{}}
\makeatother

\begin{document}

\title{Multicriticality in the Generalized Schwinger Model}

\author{Federico Ambrosino}
\affiliation{Perimeter Institute for Theoretical Physics, Waterloo, Ontario N2L 2Y5, Canada}

\author{Bernardo Zan}
\affiliation{Dipartimento di Fisica, Universit\`a di Genova and INFN, Sezione di Genova, Via Dodecaneso 33, 16146, Genoa, Italy}

\begin{abstract}
We consider the generalized Schwinger model. This theory is obtained by coupling the field strength to an auxiliary scalar field $B$ and allowing generic potentials $V(B)$. We study the properties and phases of the model, focusing on the case where the potential is quartic. We investigate the model using both semiclassical and numerical methods, and find a rich phase diagram in the space of couplings. A region with unbroken charge conjugation is separated from one where the symmetry is spontaneously broken by lines of first- and second-order phase transitions. At the special point where these two lines meet, we find a fixed point in the tricritical Ising universality class. We also observe infinitely many lines of first order phase transitions ending at points of the Ising universality class.
\end{abstract}

\maketitle

\section{Introduction}

Two-dimensional gauge theories continue to provide a valuable setting for understanding strongly coupled quantum field theories. They exhibit many of the qualitative phenomena familiar from higher-dimensional gauge theories, including confinement, while remaining far more accessible to analytic and numerical methods. A paradigmatic example is the Schwinger model \cite{Schwinger:1962tp,Coleman:1975pw,Coleman:1976uz}, namely QED in \(1+1\) dimensions. Historically, it provided the first example of a gauge theory whose infrared spectrum is gapped. Today, the Schwinger model and its generalizations remain an important laboratory for ideas ranging from high-energy theory, where it sheds light on generalized symmetries and anomalies \cite{Misumi:2019dwq, Dempsey:2022nys} and the phases of gauge theories \cite{Dempsey:2023gib,Mouland:2025dct,Cuomo:2026kmy}, to many-body physics, where it is a great tool to test numerical algorithms \cite{Byrnes:2002gj,Banuls:2013jaa,Shimizu:2014fsa,ArguelloCruz:2025zuq,Fujii:2024VUMPS}, and quantum simulations \cite{Hauke:2013jga,Martinez:2016yna,Banuls:2019bmf,Halimeh:2023lid}.

The Schwinger model is defined by
\begin{equation}\label{eq:Schwinger}
\begin{aligned}
\mathcal L ={}& -\frac{1}{4e^2} F^{\mu\nu}F_{\mu\nu}
 +\frac{\theta}{4\pi}\epsilon^{\mu\nu}F_{\mu\nu}\\
&\hspace{3em}+\overline\psi\gamma^\mu(i\partial_\mu-A_\mu)\psi-m\overline\psi\psi\,.
\end{aligned}
\end{equation}
For vanishing fermion mass $m=0$, the theory can be exactly solved through bosonization and it can be shown to be the theory of a free massive scalar, with mass $M=e/{\sqrt{\pi}}$, and independent of $\theta$. For non-zero fermion mass, the theory is not exactly solvable anymore, but can be studied numerically. We will focus on the case $\theta=0,\pi$, where the theory has a $\mathbb{Z}_2$ charge conjugation. For $\theta=0$, numerical methods allow us to establish the following phase diagram: for $m \ge m_{\rm crit}$, with $m_{\rm crit}<0$, the theory is in a $\mathbb{Z}_2$ preserving vacuum; for $m < m_{\rm crit}$, the theory is in a phase where the $\mathbb{Z}_2$ symmetry is spontaneously broken \cite{Coleman:1976uz,Byrnes:2002gj}. The value of $m_{\rm crit}$ can be studied to very high accuracy using numerical methods \cite{ArguelloCruz:2025zuq,Fujii:2024VUMPS}. These two phases are separated by a continuous phase transition, which is of the Ising universality class. By the Schwinger anomaly, for $\theta=\pi$, the phase diagram can be obtained from the $\theta=0$ diagram by replacing $m \to -m$.

The phase diagram becomes more interesting as we add more couplings or matter to the model: for example the Schwinger model with two fermions shows a fascinating phase diagram with lines of Ising CFTs, as well as a $c=1$ CFT with $SU(2)$ symmetry in the massless case, and BKT-like behavior for some range of parameters \cite{Dempsey:2023gib}.

In this work, we consider the Generalized Schwinger model \cite{Douglas:1994pq} obtained  by coupling an auxiliary scalar field with a generic potential to the gauge field \cite{Witten:1991we,Ganor:1994bq,Khorrami:1996gYM}. This model is described by the  Lagrangian
\begin{equation}\label{eq:genschwinlag}
\begin{aligned}
\mathcal L ={}& \frac12 B\epsilon^{\mu\nu}F_{\mu\nu}-e^2V(B)
 +\frac{\theta}{4\pi}\epsilon^{\mu\nu}F_{\mu\nu}\\
&\hspace{2em}+\overline\psi\gamma^\mu(i\partial_\mu-A_\mu)\psi-m\overline\psi\psi
\end{aligned}
\end{equation}
where $B$ is the  auxiliary  field, and $V$ is an arbitrary potential thereof.  Choosing a quadratic potential for $B$ gives the Lagrangian \eqref{eq:Schwinger} once $B$ is integrated out. Different choices of $V$ will instead give different theories. We are interested in the phase diagram of such theories. It is easy to expect that more complicated potentials, that depend on more parameters, may give rise to less generic (e.g. tricritical, tetracritical, \ldots) multicritical fixed points.\footnote{The Ising tricritical point has two relevant singlet operators, therefore it can be reached if a theory has two couplings that can be tuned to their critical values; the tetracritical fixed point has three relevant singlet operators, and requires three couplings to be tuned; and so on. The exception is if a symmetry larger than $\mathbb{Z}_2$ is imposed on the lattice Hamiltonian; then less tuning is required. See e.g. \cite{OBrien:2017wmx,Gabai:2024puk,Gabai:2024qum} for examples of such cases.}

The phase diagram of this theory in the  space of the possible potentials $V(B)$ is clearly very rich and interesting. An illustration of this was given in \cite{Ambrosino:2024prz}. There the authors studied the non-abelian version of the theory  \eqref{eq:genschwinlag} for a large number of colors $N_c\to \infty$. In that limit the generalized QCD theory is solvable in a way that is akin to the 't Hooft model \cite{tHooft:1974pnl,Ambrosino:2023dik, Litvinov:2024riz}. The analysis of phases for Generalized QCD in \cite{Ambrosino:2024prz} suggests the existence of (multi)critical points in the (complex) space of the couplings of the potential $V(B)$. Yet, the question of determining the fate of such critical points at a finite number of colors remained open there. One possibility could be that such criticalities are an artifact of the large $N_c$ expansion.\footnote{This was already deemed unlikely in \cite{Ambrosino:2024prz}. Indeed, studying another limit of the theory obtained by coupling  $SU(2)$ to quarks valued in large isospin representation of the gauge group, leads to exactly the same universality class.}
From this perspective, studying multicriticality in the parametrically opposite regime of the Schwinger $U(1)$ case can shed light on this question, and serves as one of the driving motivations of this work. 

For definiteness, let us consider the generalized Schwinger model with the quartic potential\footnote{ In \cite{Ambrosino:2024prz}, the potential  $V(B) = B^2 + i|g| B^3$ was considered. This gives a real spectrum of bound states as the corresponding theory is $\mathcal{PT}$-symmetric. We could have also performed the analysis of the cubic potential here, but we found it advantageous to work directly with a real Hamiltonian to simulate it on the lattice.}
\begin{equation}
    V(B)=\frac{1}{2}\left( B^4 +\tau B^2\right)
\end{equation}
with $\tau$ a dimensionless parameter. We will study the phase diagram of this model at $\theta=0,\pi$ and establish that for a special value of $\tau$ and $m$ the IR of the theory is described by a tricritical fixed point. This confirms that the existence of such critical points is an intrinsic property of the generalized theory, as suggested in \cite{Douglas:1994pq, Ambrosino:2023dik}.

Already for the case of the  simple quartic potential, the phase diagram of the theory in the plane of the two couplings $\tau$ and $m$ is extremely rich. 
We will argue, using a semiclassical approach, that the phase diagram is as follows: a region where $\mathbb{Z}_2$ is preserved is separated from a region where $\mathbb{Z}_2$ is broken by a line of first order phase transition and a line of second order phase transitions. This hints at the presence of a tricritical fixed point where the two lines meet. In the $\mathbb{Z}_2$ broken phase, we find an infinite number of first order phase transition lines, which end in critical points of the Ising universality class.

We confirm this analysis by a detailed numerical study of the model. Upon discretizing it on the lattice, we determine the universality class of the tricritical fixed point to be the same as that of tricritical Ising with central charge $c=7/10$. We mention that tricritical fixed points were already found in deformations of the lattice Schwinger model \cite{Cheng:2025rwz}, where the electric field was truncated. Here we see such a critical point in a different deformation of the theory both in the continuum and the lattice.

\section{Generalized Schwinger model: continuum formulation}
We consider the Lagrangian
\begin{equation}\label{eq:L}
\begin{aligned}
\mathcal L ={}& \frac12 B\epsilon^{\mu\nu}F_{\mu\nu}-e^2V(B)
 +\frac{\theta}{4\pi}\epsilon^{\mu\nu}F_{\mu\nu}\\
&\hspace{1em}+\overline\psi\gamma^\mu(i\partial_\mu-A_\mu)\psi-m\overline\psi\psi\\
={}& BF_{01}-e^2V(B)+\frac{\theta}{2\pi}F_{01}\\
&\hspace{1em}+\overline\psi\gamma^\mu(i\partial_\mu-A_\mu)\psi-m\overline\psi\psi
\end{aligned}
\end{equation}
where $F_{01}$ is the only independent component of $F_{\mu \nu}$ in two dimensions. We will choose the potential to be even, $V(-B)=V(B)$. Since $B$ appears in the action without derivatives, it is an auxiliary field and can be integrated out. Its equation of motion depends on the choice of the potential $V$.
The simplest choice is the quadratic potential $V(B)=B^2/2$, for which $B=F_{01}/e^2$, and the standard Schwinger-model Lagrangian is recovered. For a generic potential, integrating out $B$ gives the Lagrangian
\begin{equation}\label{eq:Lf}
\begin{aligned}
\calL ={}& e^2V^*\left(\frac{F_{01}}{e^2}\right)+\frac{\theta}{2\pi}F_{01}+\overline\psi\gamma^\mu(i\partial_\mu-A_\mu)\psi-m\overline\psi\psi
\end{aligned}
\end{equation}
where $V^*$ is the Legendre transform of $V$.\footnote{This needs to be done carefully if $V$ is not convex, where we consider different branches where $V'$ is invertible.} One may therefore choose the potential such that $V^*$ contains, in addition to the usual $F_{01}^2$ term, higher-order terms such as $F_{01}^4$, $F_{01}^6$, and so on. From this perspective, the theory is a generalization of the Schwinger model obtained by adding further couplings.

Rather than proceeding by integrating out $B$, it will be more convenient to integrate out $F_{01}$. As usual in two dimensions, the fermionic theory can be bosonized by expressing the conserved current in terms of a periodic scalar $\phi\sim\phi+2\pi$ \cite{Coleman:1974bu,Mandelstam:1975hb, Douglas:1994pq}
\begin{equation}
    j^\mu = \bar \psi \gamma^\mu \psi = \frac{1}{2\pi} \epsilon^{\nu \mu}\p_\nu \phi\,.
\end{equation}
In terms of these variables, the theory becomes
\begin{equation}\label{eq:L_before}
\begin{aligned}
\calL ={}& BF_{01}-e^2V(B)+\frac{\theta}{2\pi}F_{01}\\
&+\frac{1}{8\pi}(\p\phi)^2+\frac{1}{2\pi}\phi F_{01}+\Lambda m\cos\phi
\end{aligned}
\end{equation}
with $\Lambda$ a convention dependent energy scale. Since $F_{01}$ appears only linearly in the action, it can now be integrated out. This simply sets
\begin{equation}
    B = -\frac{\phi+\theta}{2\pi}\,.
\end{equation}
We then obtain the Lagrangian \cite{Douglas:1994pq}
\begin{equation}\label{eq:Lsingen}
\calL ={} \frac{1}{8\pi}(\p\phi)^2+\Lambda m\cos\phi-e^2V\left(\frac{\phi+\theta}{2\pi}\right)
\end{equation}
where we've used the fact that the potential is even.\footnote{We are ignoring the fact that, due to $\phi$ periodicity, we should sum over sectors with different fluxes, see e.g. Appendix A of \cite{Mouland:2025dct}. From the point of view of the IR theory this subtlety is not very important, and we can simply work with \eqref{eq:Lsingen}.}
Once again, choosing a quadratic potential reproduces the massive Schwinger model. In the following, however, we will consider more general potentials. Upon choosing the potential to be a polynomial $V(B) = \sum_n c_n B^{2n}$, and setting $m=0$, one recognizes \eqref{eq:Lsingen} to reduce exactly to a Landau-Ginzburg (LG) Lagrangian for the single scalar field $\phi$ \cite{Douglas:1994pq}. This famously \cite{Zamolodchikov:1986db} provides  a Lagrangian description of the unitary minimal models $M(p,p+1)$ upon tuning of the LG couplings, describing multicritical CFTs. 
Hence, already this simple analysis based on bosonization suggests that multicritical points might exist also for the generalized Schwinger model.

\subsection{The Hamiltonian}
Let us now return to the Lagrangian \eqref{eq:L}, before integrating out either $B$ or $F_{01}$. Since our ultimate goal is to formulate the theory on the lattice, and Hamiltonian methods are particularly powerful in $1+1$ dimensions, we begin by deriving the Hamiltonian of the system.

The momentum conjugate to \(A_1\) is\footnote{By electric field, one normally means two things: either the only independent component of the field strength $F_{01}$, or the conjugate momentum of $A_1$. For the usual Schwinger model, the two quantities agree up to factors of $e$. For a generic potential $V$, the two do not; in this work, we commit to the second choice.}
\begin{equation}
E \equiv \frac{\partial \mathcal L}{\partial \dot A_1}
=
B+\frac{\theta}{2\pi}\,.
\end{equation}

We obtain the Hamiltonian density
\begin{equation}
\begin{aligned}
\mathcal H ={}& e^2V\left(E-\frac{\theta}{2\pi}\right)+E\,\partial_1 A_0+A_0j^0\\
&+\overline\psi\gamma^1(-i\partial_1+A_1)\psi+m\overline\psi\psi\,.
\end{aligned}
\end{equation}

Integrating the \(E\,\partial_1 A_0\) term by parts and dropping boundary terms, we obtain
\begin{equation}\label{eq:H+gauss}
\begin{aligned}
\mathcal H ={}& e^2V\left(E-\frac{\theta}{2\pi}\right)-A_0(\partial_1E-j^0)\\
&+\overline\psi\gamma^1(-i\partial_1+A_1)\psi+m\overline\psi\psi\,.
\end{aligned}
\end{equation}
Since \(A_0\) appears without derivatives, its equation of motion imposes Gauss law,
\begin{equation}
\partial_1E=j^0\,,
\end{equation}
which states that the electric field changes as we cross a fermion.

Since \(A_\mu\) is a \(U(1)\) gauge field, its conjugate momentum is quantized. With the present normalization this gives
\begin{equation}
E \in \mathbb Z\,,
\qquad
B=E-\frac{\theta}{2\pi}\,.
\end{equation}
The same result would follow if we were to compute the Hamiltonian from the Lagrangian \eqref{eq:Lf} once we've integrated out $B$. Indeed,
\begin{equation}
E=\frac{\partial\mathcal L_{\rm eff}}{\partial\dot A_1}
=(V^*)'\left(\frac{F_{01}}{e^2}\right)+\frac{\theta}{2\pi}=B+\frac{\theta}{2\pi}\,,
\end{equation}
where we used that the derivatives of \(V\) and \(V^*\) are inverse functions. Passing to the Hamiltonian is then equivalent to performing the inverse Legendre transform, giving again \eqref{eq:H+gauss}.

All in all, once we impose the Gauss law, the Hamiltonian is simply
\begin{equation}\label{eq:H_cont}
\mathcal H ={} e^2V\left(E-\frac{\theta}{2\pi}\right)+\overline\psi\gamma^1(-i\partial_1+A_1)\psi+m\overline\psi\psi
\end{equation}
with $E \in \mathbb{Z}$.
\subsubsection{The Schwinger anomaly}
It's a well-known fact that, for the massless Schwinger model, the chiral symmetry is anomalous. This can be seen in several ways, such as regularizing the measure \cite{Fujikawa:1979ay} or bosonizing the theory \cite{Lowenstein:1971fc}. Yet another way is to check the commutation relations of operators in the Hamiltonian \cite{Manton:1985jm}, see also \cite{Dempsey:2022nys} which we follow.

We first build the naive conserved current
\begin{equation}
    j^{5 \mu} = \bar{\psi}\gamma^5 \gamma^\mu \psi
\end{equation}
Using the equal time commutation relations
\begin{equation}
    \begin{aligned}
       &\{\psi(t,x), \psi(t,y)^\dag \} = \delta(x-y)\\
       &[A_1(t,x), E(t,y)] = i \delta(x-y)
    \end{aligned}
\end{equation}
we obtain the commutation relations
of $j^{50}$
\begin{equation}
\begin{aligned}
    &[j^{50}(t,x),\psi(t,y)] = \gamma^5 \psi(y) \delta(x-y)\\
    &[j^{50}(t,x), \psi(t,y)^\dag] = - \psi(y)^\dag \gamma^5 \delta(x-y)\\
    &[j^{50}(t,x),A_1(t,y)] = 0
\end{aligned}
\end{equation}

The main point is that, while the naive conserved current
would commute with the electric field, we actually need to regularize this current. One option is to do so by point splitting. When we point split, we need to ensure that the current is still gauge invariant. We achieve this by inserting a gauge field string between the two fermionic operators.
\begin{equation}
    j^{50}_{\rm reg} \equiv -\psi^\dag(x+\eps) \gamma^5 e^{-i \int_{x}^{x+\eps} dy\, A_1(y)} \psi(x)
\end{equation}
At this point, it's easy to see that the commutation relation of the current with the electric field is spoiled by the presence of $A_1$. The upshot is that \cite{Manton:1985jm,Dempsey:2022nys}, in the $\eps \to 0$ limit, the naive conserved charge
\begin{equation}
    Q^5 = \int dx\, j_{\rm reg}^{50}(x)
\end{equation}
doesn't commute with the electric field,
\begin{equation}
    [Q_5, E(x)] = -\frac{i}{\pi}\,.
\end{equation}

This means that the action of a chiral transformation
\begin{equation}
    \mathcal V_{\alpha} = e^{i \alpha Q_5}
\end{equation}
changes the electric field by
\begin{equation}
   \mathcal  V_{\alpha} E (x) \mathcal V_{\alpha}^{-1} = E(x)+\frac{\alpha}{\pi}\,.
\end{equation}
It follows that
\begin{equation}
\mathcal V_\alpha\left(E(x)-\frac{\theta}{2\pi}\right)^n\mathcal V_\alpha^{-1}=\left(E(x)-\frac{\theta-2\alpha}{2\pi}\right)^n\,,
\end{equation}
and for a generic potential $V$
\begin{equation}
\mathcal V_\alpha V\left(E-\frac{\theta}{2\pi}\right)\mathcal V_\alpha^{-1}=V\left(E-\frac{\theta-2\alpha}{2\pi}\right)\,.
\end{equation}
This means that, in the same way as in the usual Schwinger model, a chiral transformation of the massless generalized Schwinger model changes the $\theta$ angle
\begin{equation}\label{eq:anomaly_cont}
\mathcal V_\alpha H(m=0,\theta)\mathcal V_\alpha^{-1}=H(m=0,\theta-2\alpha)\,.
\end{equation}

\subsection{The Generalized Schwinger model with quartic potential}
The choice of a quadratic potential gives the usual Schwinger model, which has been extensively studied. We instead choose the following potential
\begin{equation} \label{eq:V}
    V(B)=\frac{1}{2} \left( \tau B^2 + B^4\right)
\end{equation}
where $\tau$ is a dimensionless coupling which can be both positive and negative. We will be interested in understanding the low energy behavior of this model as we change the value of $\tau$ and of the fermion mass $m$. 

Any choice of $V(B)$ even, such as \eqref{eq:V}, means that the system has a $\mathbb{Z}_2$ charge conjugation symmetry $C$ when $\theta=0,\pi$. This symmetry acts as
\begin{equation}
   C:\begin{aligned}
     \quad \psi &\to \gamma^5 \psi^{*}
        \\
         A_\mu &\to - A_\mu\\
         B &\to -B
\end{aligned}
\end{equation}

Charge conjugation flips the sign of the $\theta$ angle, and therefore is only a symmetry when $\theta = 0,\pi$, given the $2\pi$ periodicity of $\theta$. Let us focus on the $\theta=0$ case ($\theta = \pi$ follows by changing the sign of the fermion mass). We would like to understand what the phase diagram of this theory is. When is the theory gapped? When is it gapless? When is the $\mathbb{Z}_2$ symmetry spontaneously broken?

To answer this question, we will eventually have to resort to numerical methods, given that the system is strongly coupled and analytic predictions are out of reach. We will do this by discretizing this model and placing it on the lattice. Before doing that, however, it's convenient to do a semiclassical analysis in the region of large couplings.
\subsection{Semiclassical analysis}

\paragraph{Large positive fermion mass}
When the mass of the fermion is very large and positive, $m\gg e$, the fermion decouples from the theory. Neglecting corrections of order $e/m$ and higher, we are therefore left with the Hamiltonian
\begin{equation}
\mathcal H =
e^2 \, V\left(B\right) = \frac{e^2}{2}(B^4 + \tau B^2)
\end{equation}
with $B\in \mathbb{Z}$.
The ground state is the state that minimizes this energy with $B$ constant (the Gauss law reduces to $\partial_1 E =0 $ when fermions are infinitely massive). The value of $B$ which minimizes the potential depends on $\tau$, and crucially has to be integer, because of the quantization condition. For positive $\tau$, $B=0$ minimizes $V(B)$. For negative $\tau$, $B$ can acquire non-zero values. More specifically, the ground state is
\begin{align}
    B=\begin{cases}
        0, & \tau \ge -1\\
       \pm n, & -(2n^2+2n+1)\le\tau\le-(2n^2-2n+1)
    \end{cases}
\end{align}
where $n \in 
\mathbb{Z}_{>0}$.
For $\tau>-1$, the ground state has $\braket{B}=0$, and $\mathbb{Z}_2$ is unbroken. For $\tau<-1$, instead $\braket{B}\neq0$ and the symmetry is spontaneously broken. The two phases are separated by a first order phase transition, where $B$ jumps discontinuously. Then, inside the region of broken $\mathbb{Z}_2$ there are further discontinuous transitions, where the value of $B$ jumps by $\pm 1$.

\paragraph{Large negative fermion mass}
When the mass is very large but negative instead, $-m\gg e$, a very similar reasoning goes through; the only difference is that integrating out a fermion with negative mass shifts the theta angle by $\pi$. Said in other words, we can use the chiral anomaly to make a chiral rotation which changes the sign of the mass of the fermion and generates a $\theta=\pi$ angle. Then the question is to minimize the potential $V(B)$ with $B \in \mathbb{Z} +\frac 12$.

The ground state is
\begin{align}
    B=\begin{cases}
        \pm \tfrac{1}{2}, & \tau \ge -\tfrac{5}{2}\\
       \pm (n+\tfrac{1}{2}), & -(2n^2+4n+\tfrac{5}{2})\le\tau\le-(2n^2+\tfrac{1}{2})
    \end{cases}
\end{align}
with again $n \in 
\mathbb{Z}_{>0}$. This
means that the $\mathbb{Z}_2$ symmetry is always spontaneously broken for $-m \gg e$. Due to the quantization condition of $B$, we still have discontinuous transitions at negative values of $\tau$, where $B$ jumps by $\pm 1$.

\paragraph{Large positive $\tau$}
When $\tau $ approaches infinity, $\tau \gg 1$, then we can neglect the quartic term, and recover the Schwinger model with $e_{\rm eff} = e\sqrt{\tau}$. It's known that the $\theta=0$ Schwinger model has charge conjugation unbroken for $m \ge m_{\rm crit}\sim -1/3 e_{\rm eff} $  \cite{Coleman:1976uz,Byrnes:2002gj} (see \cite{ArguelloCruz:2025zuq,Fujii:2024VUMPS} for the most precise values), and spontaneously broken for $m< m_{\rm crit}$, with the two phases separated by a second order phase transition in the Ising universality class. This means that, in the case of the generalized Schwinger model for very large $\tau$, we expect an Ising phase transition for $m \sim -1/3 \sqrt{\tau} e$ separating a phase of broken and unbroken $\mathbb{Z}_2$.

\paragraph{Large negative $\tau$}
For large and negative $\tau$ the potential $V(B)$ has two minima at $B=B_0 \equiv \pm \sqrt{|\tau|/2}$. Let's for now consider the positive branch for $B_0$. We can expand $B$ around this value, $B=B_0+\delta B$, and write the Hamiltonian as
\begin{equation}
\begin{aligned}
\mathcal H ={}& e^2\left(|\tau|\delta B^2+\sqrt{2|\tau|}\delta B^3+\frac12\delta B^4\right)\\
&+\overline\psi\gamma^1(-i\partial_1+A_1)\psi+m\overline\psi\psi\,.
\end{aligned}
\end{equation}
where we are ignoring constant terms. For $-\tau \gg 1$, we can neglect the cubic and quartic terms in $\delta B$, and we're left with a Hamiltonian that looks very similar to that of the usual Schwinger model. However, we have to be careful about the quantization condition. Given that we have $\theta=0$, $B$ is an integer. This means that $\delta B \in \mathbb{Z}-\sqrt{\frac{|\tau|}{2}}$, so we are landing on the Schwinger model with $\theta_{\rm eff} = 2\pi \sqrt{\frac{|\tau|}{2}}$. If $\theta_{\rm eff} \neq 0,\pi$, we expect to have a gapped theory for any value of the mass.

Things can be different if $\theta_{\rm eff}=0,\pi$. Indeed, for these values we expect to have a richer structure in our phase diagram, with second order phase transitions as $m$ is changed. We also expect first order phase transitions as $\theta$ is changed from $0^+$ to $0^-$, if the mass is negative enough, and when $\theta$ is changed from $\pi^+$ to $\pi^-$, if the mass is positive enough. Indeed, we know that there are phase transitions at large $|m|$ as $\tau$ is changed.

Let's first consider the first order transitions that appear at large enough $m$. These appear at $\tau =-2(n^2+n+1/2)$, for $n$ a positive integer, which means
\begin{equation}
    \theta_{\rm eff} = 2\pi \left( n+\frac{1}{2}+O\left( \frac{1}{n}\right) \right)\,.
\end{equation}
This is precisely what we expect. If ${m}/{e_{\rm{eff}}}=m/(\sqrt{2|\tau|}e)$ is larger than the critical value (of roughly $1/3$), then we expect a line of first order phase transitions as $\tau$ is changed. This line ends at approximately $m/e \simeq \frac{\sqrt{2|\tau|}}{3}$.

The line of phase transitions we see at large and negative $m$ instead happens at $\tau = -2(n^2+\frac{1}{4})$, meaning
\begin{equation}
    \theta_{\rm eff} = 2\pi \left( n+O\left( \frac{1}{n}\right) \right)\,.
\end{equation}
By the same line of reasoning, we expect these lines of phase transitions to end at $m/e \simeq - \frac{\sqrt{2|\tau|}}{3}$ for $-\tau\gg 1$.

\subsection{The phase diagram}
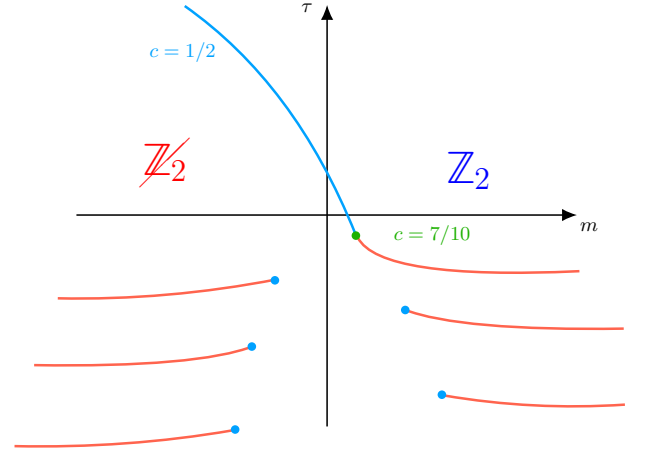
\begin{figure}[!tbp]
\centering
\resizebox{\columnwidth}{!}{\usetikzlibrary{arrows.meta}
\definecolor{bluecurve}{rgb}{0.0006,0.6336,1.0}
\definecolor{bluedot}{rgb}{0.0,0.6332,1.0}
\definecolor{redcurve}{rgb}{1.0,0.3907,0.3056}
\definecolor{greendot}{rgb}{0.113,0.6938,0.0017}

\begin{tikzpicture}[
    scale=0.7,
    axis/.style    ={black, line width=0.7pt},
    flow/.style    ={redcurve, line width=1.15pt},
    critical/.style={bluecurve, line width=1.15pt},
  ]
  \draw[axis,-{Latex[length=2.4mm,width=2.0mm]}] (9.6,0.92) -- (9.6,10.55);
  \draw[axis,-{Latex[length=2.4mm,width=2.0mm]}] (3.875,5.749) -- (15.31,5.749);
  \draw[critical] (6.349,10.523) .. controls (8.032,9.334) and (9.316,7.621) .. (10.25,5.31);
  \draw[flow] (15.363,4.467) .. controls (12.33,4.329)  and (10.643,4.584) .. (10.302,5.232);
  \draw[flow] (16.375,3.155) .. controls (13.889,3.118) and (12.218,3.263) .. (11.362,3.59);
  \draw[flow] (16.402,1.419) .. controls (15.001,1.327) and (13.626,1.397) .. (12.278,1.627);
  \draw[flow] (3.449,3.847)  .. controls (5.117,3.813)  and (6.788,3.958)  .. (8.462,4.282);
  \draw[flow] (2.909,2.321)  .. controls (5.395,2.284)  and (7.066,2.429)  .. (7.921,2.756);
  \draw[flow] (2.461,0.471)  .. controls (4.175,0.435)  and (5.834,0.555)  .. (7.439,0.831);
  \fill[greendot] (10.258,5.279) circle[radius=0.098];
  \fill[bluedot]  (8.408,4.258)  circle[radius=0.098];
  \fill[bluedot]  (7.881,2.745)  circle[radius=0.098];
  \fill[bluedot]  (7.5,0.85)  circle[radius=0.098];
  \fill[bluedot]  (11.385,3.58) circle[radius=0.098];
  \fill[bluedot]  (12.222,1.645) circle[radius=0.098];
  \node at (9.146,10.48) {$\tau$};
  \node at (15.583,5.48) {$m$};
  \node[bluecurve] at (6.3,9.47)  {$c = 1/2$};
  \node[greendot] at (12,5.283){$c = 7/10$};
  \node[red] at (5.893,7.017) {\huge $\cancel{\mathbb{Z}_2}$};
  \node[blue] at (12.829,6.768){\huge $\mathbb{Z}_2$};
\end{tikzpicture}}

    \caption{A scenario for the phase diagram of the generalized Schwinger model with a quartic coupling at $\theta=0$. A region where the $\mathbb{Z}_2$ symmetry is preserved is separated from a region where $\mathbb{Z}_2$ is spontaneously broken by a line of first order phase transitions ({\color{redcurve}red}) and a line of second order phase transitions of the Ising universality class ({\color{bluecurve}light blue}). These two lines meet at a tricritical Ising fixed point ({\color{greendot}green}). In the $\mathbb{Z}_2$ broken phase, we have some lines of first order phase transitions which end in some Ising fixed points ({\color{bluecurve}light blue}).}
    \label{fig:phase_diag}
\end{figure}

With the information we obtained using the semiclassical analysis, we can make a prediction for the phase diagram of the theory. For large enough $m$ and $\tau$, we have a region of unbroken $\mathbb{Z}_2$ symmetry. This region is bounded by a line of first order phase transitions and by a line of second order phase transitions. As $\tau$ becomes negative enough, we are in the $\mathbb{Z}_2$ broken phase, and we observe lines of first order phase transitions where the order parameter jumps discontinuously. These lines end in some Ising fixed points. A qualitative drawing of the phase diagram as suggested by the semiclassical analysis is in Figure~\ref{fig:phase_diag}.

Let's turn our attention to the region of unbroken $\mathbb{Z}_2$. The fact that this region is bounded by a line of first order phase transitions and by a line of second order phase transitions means that where these lines meet we expect to see a tricritical fixed point \cite{Landau:1937obd,Griffiths:1970zza,Blume:1971zza}, which we expect to be in the universality class of tricritical Ising. This fixed point, however, is not under control in our semiclassical analysis, and the only way to verify its existence is to discretize the model on the lattice. This is what we investigate in the remainder of the paper.

\section{Lattice implementation}

\subsection{Chiral anomaly on the lattice}
We want to discretize the following Hamiltonian by placing it on the lattice
\begin{equation}
\begin{aligned}
H=\int dx\,\biggl\{&\frac{e^2}{2}\left[\left(E-\frac{\theta}{2\pi}\right)^4
 +\tau\left(E-\frac{\theta}{2\pi}\right)^2\right]\\
&+\overline\psi\gamma^1(-i\partial_1+A_1)\psi+m\overline\psi\psi\biggr\}
\end{aligned}
\end{equation}

We will use the usual Kogut-Susskind formulation of staggered fermions \cite{Kogut:1974ag,Banks:1975gq} to place the theory on a one dimensional lattice. We remind the reader that in this formulation, the upper and lower components of the fermions are placed on even and odd lattice sites, respectively. In one spatial dimension, this completely avoids the problem of fermion doubling.

Following the usual procedure, the electric field $E$ is replaced by an operator $L_n$, with integer eigenvalues, which lives on the link $n$ between sites $n$ and $n+1$. The naive lattice Hamiltonian would then be
\begin{equation}\label{eq:Hnaive}
\begin{aligned}
H_{\rm naive}={}&\frac{e^2a}{2}\sum_{n=1}^N\biggl[\left(L_n-\frac{\theta}{2\pi}\right)^4+\tau\left(L_n-\frac{\theta}{2\pi}\right)^2\biggr]\\
&+\frac{i}{2a}\sum_{n=1}^N(\chi_n^\dag U_n\chi_{n+1}-{\rm h.c.})\\
&+ m\sum_{n=1}^N(-1)^n\chi_n^\dag\chi_n
\end{aligned}
\end{equation}
where we have the following commutation relations
\begin{equation}
    \lbrace \chi_n,\chi_m^\dag \rbrace = \delta_{nm}\,, \qquad [L_n,U_m]=\delta_{nm} U_m\,.
\end{equation}
Here we make the clarification that the lattice couplings $\tau$ and $m$ are not the same as the continuum couplings $\tau$ and $m$, as they might get renormalized when we place the theory on the lattice. Rather than writing e.g. $\tau_{\rm lat}$ for the lattice parameters, we simplify the notation by using the same symbols, with the understanding that when dealing with the lattice model we are dealing with lattice parameters. For now, we discuss the PBC Hamiltonian.

To enforce gauge invariance, we also have to discretize Gauss law. The standard choice is
\begin{equation}
    L_n-L_{n-1}=Q_n\equiv \chi_n^\dag \chi_n -\frac{1-(-1)^n}{2}\,.
\end{equation}
Requiring $L_{n+N}=L_{n}$ imposes half-filling, $\sum_k n_k = N/2$.
There is some arbitrariness in the choice of $Q_n$, the charge of site $n$, and while some works have chosen a different definition \cite{Berruto:1997jv}, we follow the convention used in the majority of the literature. In the case of the ordinary Schwinger model, it was pointed out in \cite{Dempsey:2022nys} that such choice has consequences on the definition of the lattice mass. For example, while the Schwinger anomaly of the massless model is not reproduced on the lattice in its entirety, a discrete subgroup of it is for some value of the lattice mass $ m \neq0$. Then it makes sense to identify this point as the discretization of the massless continuum theory. This means that the continuum mass and lattice mass were related by a mass shift in the usual Schwinger model.

We will repeat this analysis for the generalized Schwinger model. We will see that a mass shift would not be enough to achieve this, but we can change directly the lattice formulation of the model so that for $ m= 0$ a discrete subgroup of the Schwinger anomaly is reproduced on the lattice.

Let's consider only the terms in the Hamiltonian that include the electric field $E$, as the rest of the Hamiltonian will remain unchanged
\begin{equation}
\begin{aligned}
H={}&H_E+\frac{i}{2a}\sum_n(\chi_n^\dag U_n\chi_{n+1}-{\rm h.c.})\\
&+ m\sum_n(-1)^n\chi_n^\dag\chi_n
\end{aligned}
\end{equation}
Instead of the naive choice \eqref{eq:Hnaive}, let's consider the more general form
\begin{equation}\label{eq:HE_working}
\begin{aligned}
H_E=\frac{e^2a}{2}\sum_n\biggl[&\left(L_n-\frac{\theta}{2\pi}+\beta(-1)^n\right)^4\\
&+\tau\left(L_n-\frac{\theta}{2\pi}+\beta(-1)^n\right)^2\biggr]
\end{aligned}
\end{equation}
with $\beta$ a number that we have yet to fix. No matter the value of $\beta$, this Hamiltonian is invariant under charge conjugation, which acts on the fields as \cite{Berruto:1997jv}
\begin{equation} \label{eq:CC}
C:\begin{aligned}
        \quad L_n &\to \frac{\theta}{\pi}-L_{n+1}\\
        U_n &\to  U_{n+1}^\dagger\\
        \chi_n &\to  \chi_{n+1}^\dagger\\
\end{aligned}
\end{equation}
As required, $C$ is a symmetry only for $\theta=0,\pi$, since it leaves $L_n$ an integer.

We will now fix $\beta$ by requiring that in the case of $ m=0$ a discrete subgroup of the Schwinger anomaly is preserved. Let's consider the operator which implements a translation by one lattice site $\mathcal V$. This acts on operators as \cite{Dempsey:2022nys}
\begin{equation}
\begin{aligned}
    \mathcal V \chi_n \mathcal V^{-1} &= \chi_{n+1}\\
    \mathcal V U_n \mathcal V^{-1} &= U_{n+1}\\
    \mathcal V L_n \mathcal V^{-1} &= L_{n+1}+\frac{1+(-1)^n}{2 }\,.
\end{aligned}
\end{equation}
In the staggered fermion language, translations by one site correspond to a chiral transformation by $\pi/2$ \cite{Banks:1975gq}, which given \eqref{eq:anomaly_cont} means that we should require \cite{Dempsey:2022nys}
\begin{equation}
\mathcal V H(\theta,\tau, m=0)\mathcal V^{-1}=H(\theta-\pi,\tau, m=0)
\end{equation}
Requiring this fixes $\beta=-1/4$. If we only had a quadratic term for the electric field, then we could use the Gauss law to transform this staggered term in the Hamiltonian into a shift of the mass term, like \cite{Dempsey:2022nys} did. For higher powers of the electric field, however, using the Gauss law would turn this staggered term into a non-local fermionic interaction, which we prefer to avoid as it would slow down the computation we will carry out using tensor network techniques. Therefore, we find it more convenient to work with a staggered term for the electric field.

We will therefore work with the following Hamiltonian
\begin{equation}\label{eq:H}
\begin{aligned}
H={}&\frac{e^2a}{2}\sum_{n=1}^N\biggl[\left(L_n-\frac{\theta}{2\pi}-\frac{(-1)^n}{4}\right)^4\\
&\hspace{4em}+\tau\left(L_n-\frac{\theta}{2\pi}-\frac{(-1)^n}{4}\right)^2\biggr]\\
&+\frac{i}{2a}\sum_{n=1}^N(\chi_n^\dag U_n\chi_{n+1}-{\rm h.c.})+ m\sum_{n=1}^N(-1)^n\chi_n^\dag\chi_n\,.
\end{aligned}
\end{equation}

Different choices of the discretization of the model such as \eqref{eq:Hnaive} and \eqref{eq:H} correspond to different relations between the continuum parameters and the lattice ones.
\subsection{Integrating out the gauge field}

In order to numerically simulate the model using tensor network techniques, it will often be convenient to work with open boundary conditions. In this case, the gauge invariant Hilbert space at finite $N$ is finite dimensional, and the electric field can be completely integrated out from the Hamiltonian using the Gauss law. 
The price to pay is that charge conjugation, which acts as a one site translation, see \eqref{eq:CC}, is broken by this choice of boundary conditions.

We first pass to a bosonic formulation of the model using a Jordan-Wigner transformation
\begin{equation}
	\begin{aligned}
		n_j&\equiv \chi^\dag_j \chi_j = \frac{\sigma_j^z+{1}}2\\
		\chi_n^\dag & = i^n e^{i \pi \sum_{j<n} n_j} \sigma^+_n\\
		\chi_n & = i^{-n}e^{-i \pi \sum_{j<n} n_j} \sigma^-_n\,,
	\end{aligned}
\end{equation}
which gives
\begin{equation}\label{eq:H_bos}
\begin{aligned}
H={}&\frac{e^2a}{2}\sum_{n=1}^{N-1}\biggl[\left(L_n-\frac{\theta}{2\pi}-\frac{(-1)^n}{4}\right)^4\\
&\hspace{4em}+\tau\left(L_n-\frac{\theta}{2\pi}-\frac{(-1)^n}{4}\right)^2\biggr]\\
&+\frac{1}{2a}\sum_n(\sigma_n^+ U_n\sigma_{n+1}^-+{\rm h.c.})+\frac{ m}{2}\sum_{n=1}^N(-1)^n\sigma_n^z\,,
\end{aligned}
\end{equation}
together with the Gauss law
\begin{equation}
    L_n -L_{n-1}=Q_n=\frac{\sigma^z_n+(-1)^n}{2}\,.
\end{equation}

Now we are ready to integrate out the gauge field. With OBC, we impose the boundary conditions $L_{0}=L_{N}=0$, and then we can integrate out the gauge field by either integrating out the field from the left
\begin{equation}
    L_j = \sum_{k=1}^{j}Q_k
\end{equation}
or from the right
\begin{equation}
    L_j = -\sum_{k=j+1}^{N}Q_k\,.
\end{equation}
These two alternatives are equivalent because we are working at half-filling, $\sum_{n=1}^N Q_n = 0$. The Hamiltonian can be represented as a MPO of bond dimension $8$.

\begin{figure*}[!tp]
\centering
\begin{minipage}[c]{.44\textwidth}
\centering
  \includegraphics[width=.77\linewidth]{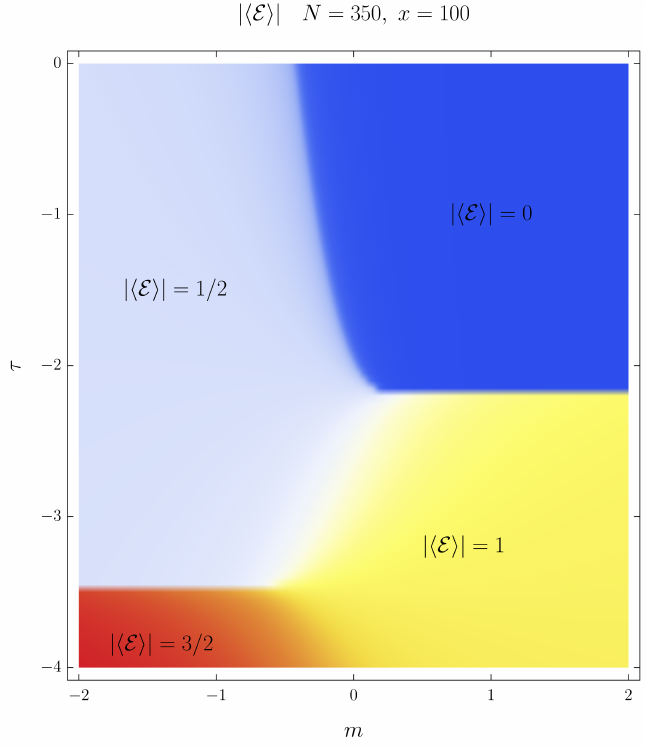}
\end{minipage}\hfill
\begin{minipage}[c]{.54\textwidth}
\centering
\includegraphics[width=\linewidth]{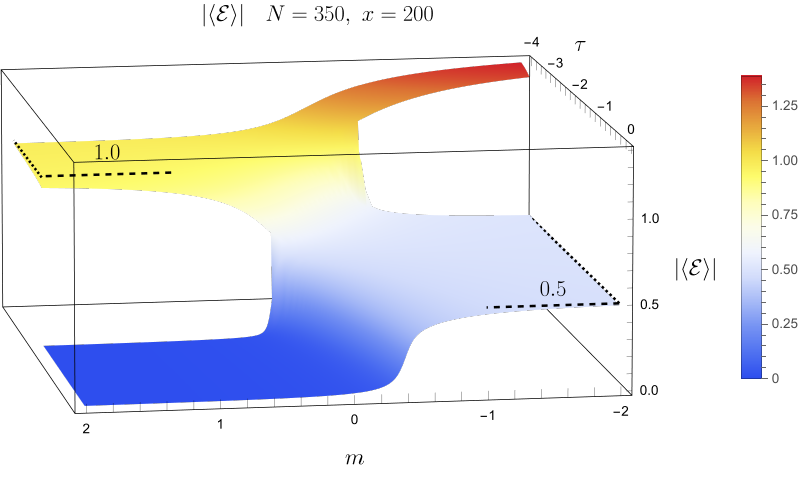}
\end{minipage}
\caption{Phase diagram obtained by DMRG with open chain boundary conditions, shown as a density plot (left) and a three-dimensional surface (right). Dashed lines in the right plot mark order-parameter values $1/2$ and $1$.}
\label{fig:phase-diagrams}
\label{fig:phases2D}
\label{fig:phases3D}
\end{figure*}

In the case of PBC, the situation is different as we cannot fully integrate out the gauge field using Gauss law. We are left with the electric field on one link, for example $L_0$. A nice basis for the Hilbert space is, for example,
\begin{equation}
    \ket{L_0; n_1, n_2,\ldots, n_N}
\end{equation}where $L_0 \in \mathbb{Z}$. This means that, even at a finite value of the lattice size $N$, the Hilbert space is infinite. In order to do numerics, we truncate the Hilbert space and only consider states which have $|L_0|\le \ell_{\rm max} $. 
The PBC Hamiltonian can be represented by a MPO of bond dimension 12.

\section{Results}
In this section we present the results obtained by DMRG \cite{White:1992zz}, using the ITensor Julia package \cite{Fishman:2020gel}.

There are two limits that we should take when doing the numerics. We should take the thermodynamic limit ($N\to\infty$) and the continuum limit $(a \to 0)$. Given that we are ultimately interested in the critical points we find in the theory, we will focus on taking the thermodynamic limit and often work at fixed lattice spacing. We will see that even in this limit the phase diagram qualitatively agrees with the continuum one discussed in Figure \ref{fig:phase_diag}. We expect that taking the continuum limit will move around the features of the phase diagram without altering them.

We will work at $\theta=0$, as the result for $\theta=\pi$ immediately follows from this by doing $m \to -m$. It is convenient to work with the dimensionless Hamiltonian $\hat H = \frac{2}{e^2 a} H$. We therefore work with the dimensionless quantity $x\equiv (e a)^{-2}$. The continuum limit corresponds to $x\to \infty$.

From here onward, we set the electric coupling constant $e$ to unity, and express all the results in units of this coupling.

\subsection{Phase diagram}
We start by describing the phase diagram of the quartic Schwinger model as a function of the two couplings $(m,\tau)$. We characterize the phases using the average electric flux in the ground state as the order parameter for $\mathbb Z_2$ symmetry breaking
\begin{equation}
\begin{gathered}
\langle\mathcal E\rangle\equiv\frac{1}{N_{\rm links}}\sum_{j=1}^{N_{\rm links}}\langle L_j\rangle
\end{gathered}
\end{equation}
where ${N_{\rm links}}=N$ for PBC and ${N_{\rm links}}=N-1$ for OBC.

Similarly to the continuum analysis, we can argue semiclassically that, for large positive values of the mass and positive enough values of $\tau$, we expect indeed a phase with $\E=0$ corresponding to the vacuum\footnote{For simplicity here we discuss the OBC case.}  
\[\ket{\uparrow\downarrow\uparrow\downarrow\cdots} \]  
with $\sigma_j^z\ket{\uparrow}_j = \ket{\uparrow}_j$ and $\sigma^z\ket{\downarrow}_j = -\ket{\downarrow}$, which has $Q_k=0$ and $L_k=0$, and therefore $\E \simeq 0$.

For large negative values of the mass, and again for $\tau$ positive enough, we expect the vacuum
\begin{equation}
    \ket{\downarrow\uparrow\downarrow\uparrow\cdots} 
\end{equation}for which $Q_k=(-1)^k$. Gauss law gives $L_j =\frac{(-1)^j-1}{2} $,
so the flux alternates between $-1$ on odd links and $0$ on even links. Averaging over the links gives
$\E=\frac{N}{2(N-1)}\to\frac12$, as $N\to\infty$.\footnote{Our choice of OBC, with $L_0=L_N=0$, explicitly breaks the $\mathbb{Z}_2$ symmetry. For the ideal staggered state considered here, these boundary conditions fix the average flux to be negative, $\langle\mathcal E\rangle\to-1/2$.}

The same analysis could be repeated as we decrease the value of $\tau$, where we again expect first order phase transitions where the order parameter changes discontinuously.

In Figure~\ref{fig:phases2D} (left) we display
the numerical results of this analysis, by plotting the expectation value of the order parameter $\E$ in the $(m,\tau)$ plane. 
 These results confirm the semiclassical analysis performed above. The dark blue region corresponds to an unbroken $\mathbb Z_2$ phase with order parameter $\E=0$. There are horizontal lines of first order phase transitions where the average electric field parameter discontinuously jumps by one unit; in the SSB region the order parameter takes integer values for large positive mass 
and half-integer ones $\mathbb Z +1/2$ for large negative mass.  
The 3D plot in Figure~\ref{fig:phases3D} (right) allows for a better visualization of the horizontal first-order phase transitions as well as the second order line.

\begin{figure*}[!tp]
\centering
\subfloat[Entanglement entropy density
\label{fig:desityS2}
  \label{fig:densityS2d}]{%
\begin{minipage}[b]{.48\textwidth}
\centering
  \includegraphics[width=.80\linewidth]{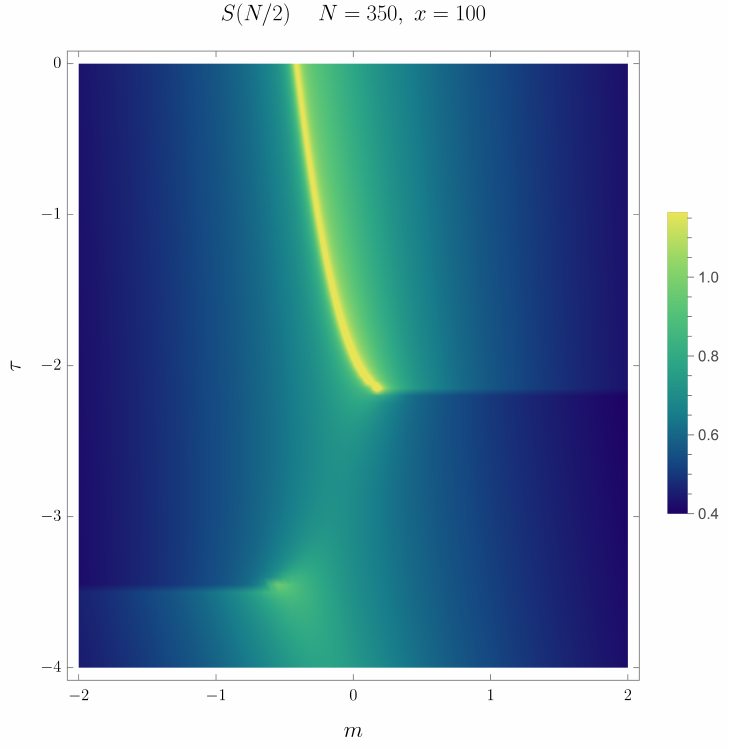}
\end{minipage}}\hfill
\subfloat[Entropy contours on the phase diagram
\label{fig:contourplot}]{%
\begin{minipage}[b]{.48\textwidth}
\centering
\includegraphics[width=.80\linewidth]{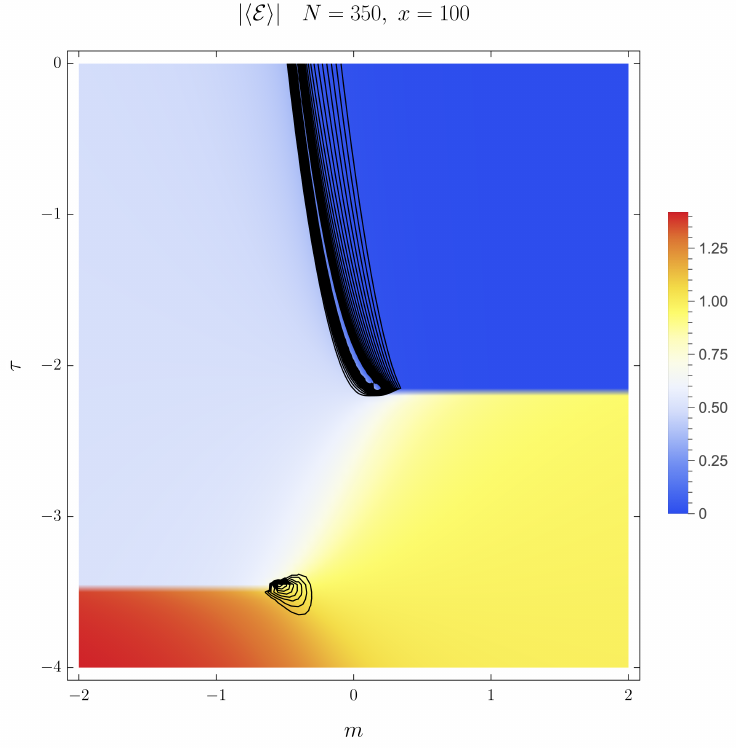}
\end{minipage}}
\caption{Entanglement entropy at the mid-chain cut $S(N/2)$ obtained by DMRG.
Panel (a) highlights a line of critical points and an isolated critical point. Panel (b) overlays entropy contours on the order-parameter phase diagram of Figure~\ref{fig:phases2D} (left).}
\label{fig:denistyS}
\end{figure*}
\begin{figure}
    \centering
    \includegraphics[width=0.9\linewidth]{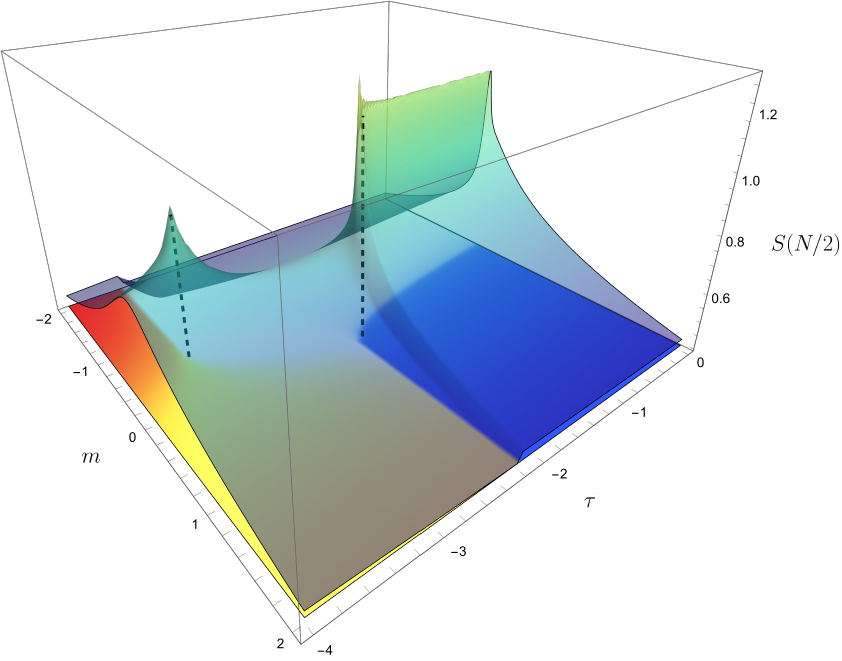}
    \caption{Three-dimensional views of the mid-chain entanglement entropy $S(N/2)$ obtained by DMRG, combined with the order-parameter density plot in the $( m,\tau)$ plane in Figure~\ref{fig:phase-diagrams}.}
    \label{fig:entropy-surfaces}
\end{figure}

\subsection{Critical points}

The phase diagram in Figure~\ref{fig:phase-diagrams} is consistent with first order phase transitions between the phases \[ \E = \mathbb Z \to \E = \mathbb Z +1   \] and
\[ \E = \mathbb Z+\frac{1}{2} \to \E = \mathbb Z +\frac{1}{2}+1   \]
along the horizontal first order lines. To confirm instead that the transition between the state with $\E=0$ and $\E=1/2$ is of second order, we perform a
preliminary   characterization by  reporting  the von Neumann entanglement entropy  measured at the central cut  $S(N/2)$ in Figures~\ref{fig:denistyS} and~\ref{fig:entropy-surfaces}.
For a bipartition after site $\ell$, we define the von Neumann entropy
\begin{equation}
 S(\ell,N)=-\operatorname{Tr}\rho_\ell\log\rho_\ell .
\end{equation}
For an open chain at a critical point, conformal invariance predicts
\begin{equation}
\label{eq:vNe}
S(\ell,N)={}\frac{c}{6}\log\left[\frac{2N}{\pi}\sin\left(\frac{\pi\ell}{N}\right)\right]+s_1+\delta S(\ell,N),
\end{equation}
where $\delta S$ contains boundary, lattice, and irrelevant-operator
corrections.  Hence, the mid-chain entropy shown in Figures~\ref{fig:denistyS} and~\ref{fig:entropy-surfaces} provides a useful qualitative
benchmark of the phase transition. The entropy illustrates critical behavior of the second order type  at the phase transition between the unbroken and broken phases with $\E =1/2$ at large $ m$, as expected from the semiclassical analysis. The entropy also confirms that the transition between the  $\E= 3/2$ and $\E =1/2$ phases is of first order, as expected. 

The contour plot in Figure~\ref{fig:contourplot} and the entropy surface in Figure~\ref{fig:entropy-surfaces}, both overlaid on the order-parameter phase diagram, illustrate how the peak for $S(N/2)$ at fixed $\tau$ coincides exactly with the critical mass at the second order line. The critical point at the intersection of the second- and first-order lines, corresponding to the expected tricritical point, is also clearly visible.

\subsubsection{Central charge at critical points}
\begin{figure}[!tbp]
\centering
  \includegraphics[width=\linewidth]{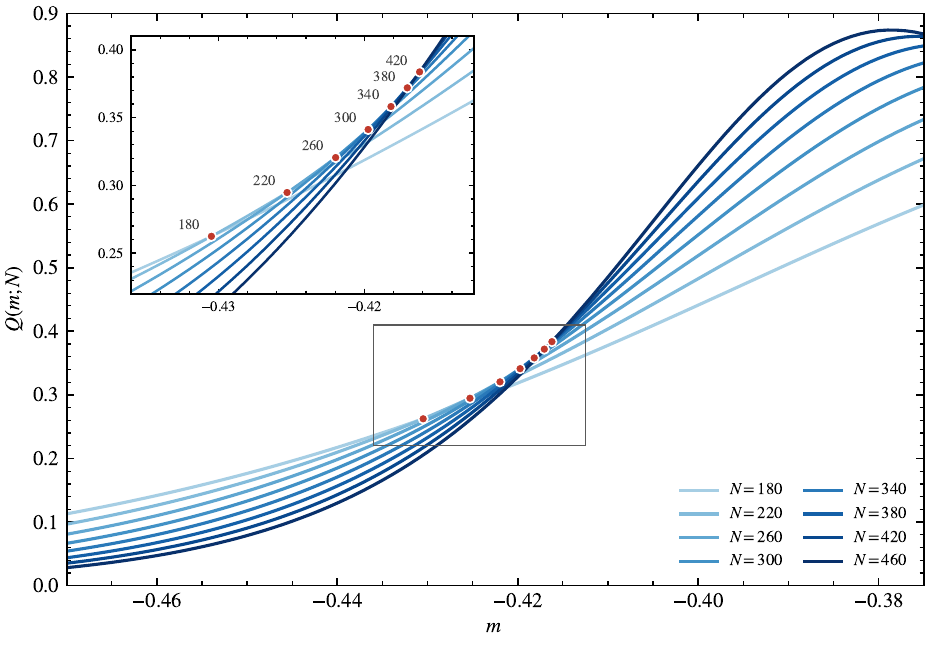}
\caption{Computation of $Q(m,N)$ for different values of $N$ in the case of $ \tau = 0$ and $x=100$.}
\label{fig:tau0}
\end{figure}

\begin{figure*}[!tp]
\centering
\includegraphics[width=.45\textwidth]{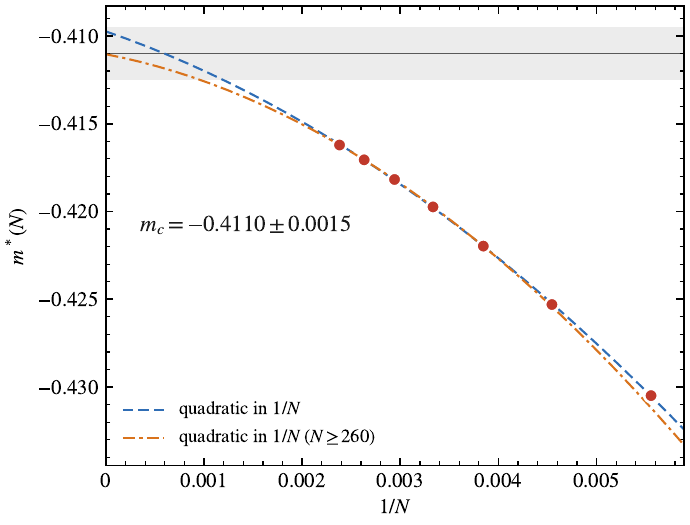}\hfill
\includegraphics[width=.45\textwidth]{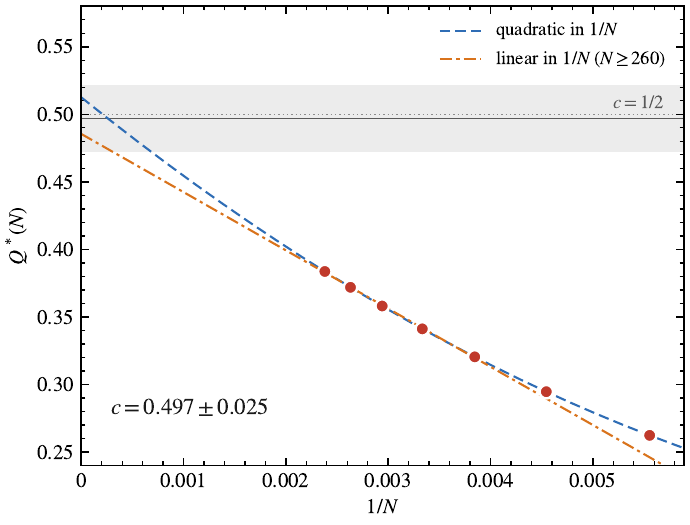}
\caption{Extrapolated $N \to \infty$ values for the critical mass and for $Q(N,m^*_N)$ at $x=100$, $\tau=0$, which gives the value of the central charge.}
\label{fig:tau0_c}
\end{figure*}
We now proceed by analyzing in detail the $2^{\rm nd}$ order phase transitions. We would like to use finite size scaling of the entropy and energy to measure the central charge of the CFTs.

We will proceed in two ways: the first method will study the entanglement entropy of the system with OBC; the second will study the energy levels of the system with PBC. Each of the two methods has its strengths: with DMRG, we can reach larger system sizes in the case of OBC; however, PBC systems are free from boundary effects.

The first method consists in studying the entanglement entropy for OBC. As explained in \cite[Sec IV]{Campostrini:2014lta}, we consider the system with $N$ spins, then split it into a left interval of length $w N$ and a right interval of length $(1-w) N$, and we compute the bipartite entanglement entropy $S(N,w,m)$.\footnote{The entanglement entropy depends on $x$ and $\tau$ as well, but in this computation we keep them fixed, so in order to simplify the notation we don't write the dependence of $S$ on them.} Then, for two choices of splitting, we define the quantity
\begin{equation}
Q(N,w_1,w_2, m)=6\left(\frac{S(N,w_1, m)-S(N,w_2, m)}
{\log\sin\pi w_1-\log\sin\pi w_2}\right)
\end{equation}
At the critical point ${m}={m}^*$, this quantity becomes independent of $w_1,w_2$ and its value for large system sizes becomes the central charge as clear from the definition \eqref{eq:vNe}
\begin{equation}
    \lim_{N\to \infty} Q(N,w_1,w_2,{m}^*) = c.
\end{equation}

We start by studying the line of second order phase transitions, which we expect to be of the Ising type. We fix the value of $x$ and $\tau$, and scan over $ m$, and compute $Q(N,{m})=Q(N,1/2,1/4,{m})$ as a function of the latter. We compute the functions $Q(N,m)$ for different values of $N$. In the limit of large $N$, the lines $Q(N,m)$ for different values of $N$ cross at the critical point. For two subsequent values of the system size $N$ and $N+\delta N$, we define ${m}^*_{N}$ as the value of ${m}$ for which $Q(N,{m}^*_{N})=Q(N+\delta N,{m}^*_N)$. Then the critical mass is given by
\begin{equation}\label{eq:m_star_limit}
    {m}^* = \lim_{N\to \infty} {m}^*_N
\end{equation}
and the central charge by
\begin{equation}
    c = \lim_{N\to\infty} Q(N,{m}^*_N)
\end{equation}

In our case, we choose $\delta N=40$, and report the results for the case of $\tau = 0$ and $x=100$ in Figure~\ref{fig:tau0} and \ref{fig:tau0_c}. Let us remark that the point $\tau=0$ is by no means a special point (corresponding to $V(B)= B^4/2$); the usual Schwinger model corresponds instead to $\tau \to \infty$. 

Most importantly, we see a critical point with central charge $c=0.497\pm 0.025$, which as expected is compatible with the Ising central charge $c=1/2$.
While the location of the critical point is not universal and depends on $x$ and other microscopic details of the model (e.g.\ the absence of other $B^n$ in the UV Lagrangian), their critical exponents are universal upon retuning to criticality. The fact that  we have worked at fixed $x$ in this analysis is not crucial.

We repeat the same computation for different values of $\tau$ and $x$, finding similar results. In the case $\tau=-0.9$ and $x=200$, we find $c=0.47 \pm 0.04$. This corroborates the idea that there is a region of parameter space where the broken and the unbroken regions are separated by a line of Ising CFTs.

\begin{figure}
    \centering
\includegraphics[width=0.9\linewidth]{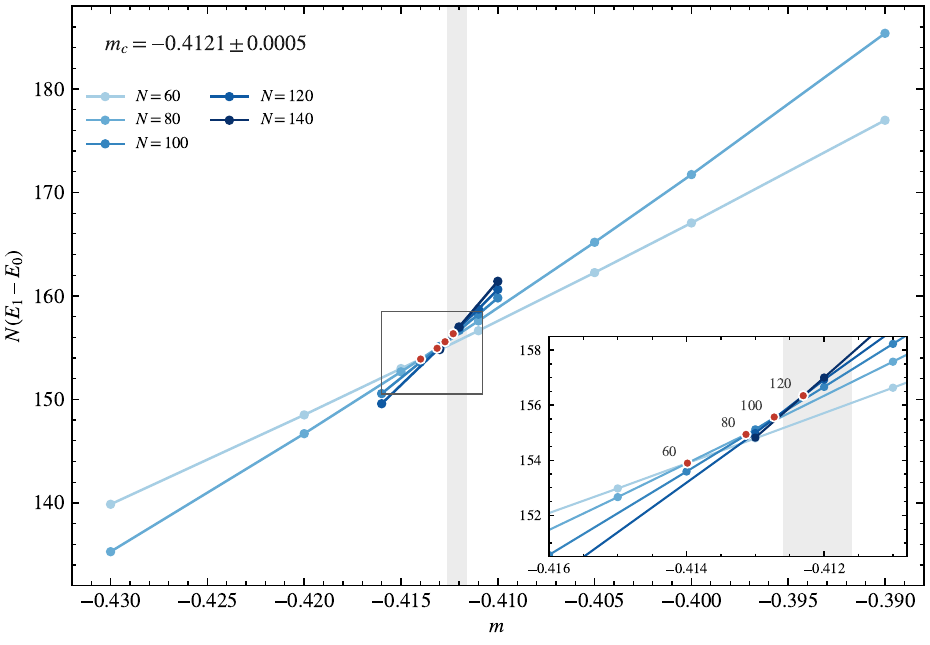}
    \caption{Rescaled energy gap as a function of the mass for the PBC chain at $x=100$, $\tau=0$. The gaps cross at the critical mass. These plots were obtained with a bond dimension up to $\chi=800$ with $\ell_{\rm max}=2$.}
    \label{fig:PBC_mc}
\end{figure}

\begin{figure}
    \centering
    \includegraphics[width=0.9\linewidth]{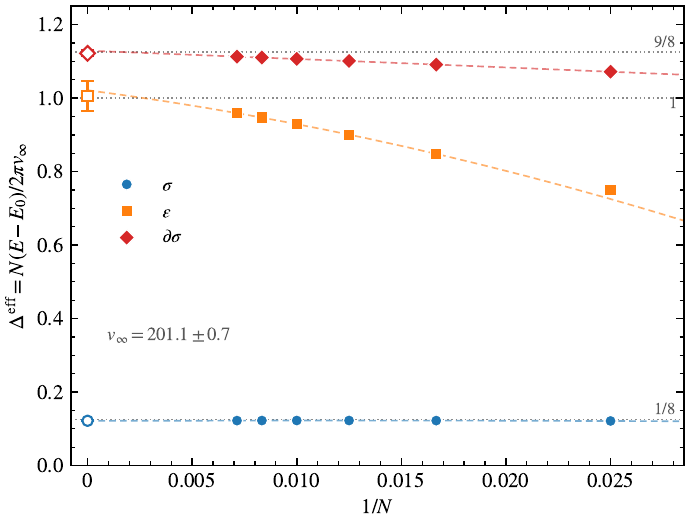}
    \caption{Spectrum at the critical point for the PBC chain, allowing a determination of $v$.}
    \label{fig:PBC_spectrum}
\end{figure}

\begin{figure}
    \centering
    \includegraphics[width=0.9\linewidth]{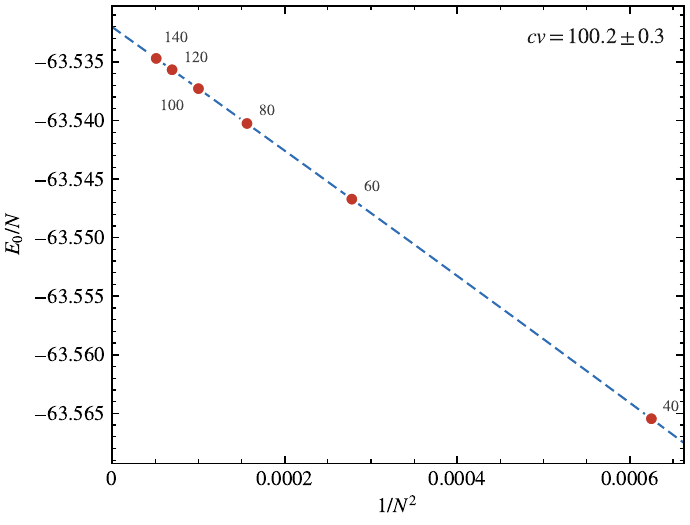}
    \caption{Convergence of $E_0$ as a function of $N$, giving the combination $cv$ thanks to \eqref{eq:cv}.}
    \label{fig:PBC_c}
\end{figure}

We can study the system with PBC as well. Due to the slower convergence when doing DMRG with PBC, we are limited to smaller values of $N$, but we don't have to worry about boundary effects. The idea is that, at the critical point, the rate of convergence of the ground state energy depends on the central charge $c$ \cite{Bloete:1986qm,Affleck:1986bv}\footnote{It is important to compute this as the couplings are kept fixed, and do not drift with $N$; i.e. one should use $m^*$ rather than $m^*_N$, see \eqref{eq:m_star_limit}. Using $m^*_N$ would introduce further $1/N$ effects that would pollute the reading of $c v$.}
\begin{equation} \label{eq:cv}
    E_0(N) = N e_\infty - \frac{\pi c v}{6 N}+\ldots
\end{equation}
where $v$ is the speed of light. 
Since the energies refer to the dimensionless Hamiltonian $\widehat H=2H/(e^2a)$, and lengths are measured in lattice sites, the velocity $v$ is therefore related to the physical propagation speed $u$ by $v=2xu$. In the Lorentz-invariant continuum theory, the normalization adopted above gives $u=1$. At finite lattice spacing, however, the effective velocity must be determined from the spectrum. 
This itself can be fixed by invoking the state operator correspondence, relating the energies on the cylinder to the scaling dimensions of CFT operators
\begin{equation}
    E_i - E_0 = \frac{2\pi v}{N} \Delta_i+\ldots
\end{equation}
We determine $v$ by requiring the scaling-dimension difference between a primary and its first
derivative descendant to be one.

The first thing we need to do is to find the critical mass, which we obtain by plotting the energy gap as a function of $N$ and looking for scale invariant points. As an example, we will again consider $\tau=0$, $x=100$. The crossing plots can be found in Figure~\ref{fig:PBC_mc}, and they give the estimate $m_c = -0.4121(5)$ which agrees with the OBC computation.

At this point we then compute the low-lying spectrum, see Figure~\ref{fig:PBC_spectrum}. By requiring the interval between the operator $\sigma$ and its descendant $\partial \sigma$ to be $1$, we fix the value of the speed of light $v$. Finally, looking at the
convergence of the ground state energy we find $c=0.497(3)$, see Figure~\ref{fig:PBC_c}, again compatible with the OBC computations.

Both the OBC and the PBC computations show that the line of second order phase transitions separating the ordered and disordered phase in the upper part of the phase diagram is a $c=1/2$ Ising line.

\subsubsection{Critical points at the end of first order phase transitions}
The phase diagram also predicts lines of first order phase transitions which end. Here we study their endpoints and check that they also are Ising CFTs.

For concreteness, we will consider the first order phase transition between $\E=1/2$ and $\E=3/2$. Across this first-order transition, the order parameter $|\langle\mathcal E\rangle|$ jumps discontinuously between the two coexisting phases. This jump decreases and eventually vanishes as we approach the Ising fixed points at the end of the first order phase transition line.

For this computation we again resort to PBC, the reason being that, 
as usual when working with first order phase transitions, we find it convenient to slightly break the global symmetry; in this way we select one of the two otherwise degenerate ground states in the broken phase. In the OBC case the boundary conditions already break the symmetry, but this effect is large and cannot be tuned. In the PBC case, instead, we can do this by adding an external field
\begin{equation}
    H \to H +h \sum_{i=1}^N L_i
\end{equation}
whose coefficient $h$ can be tuned.

\begin{figure}
    \centering
    \includegraphics[width=0.9\linewidth]{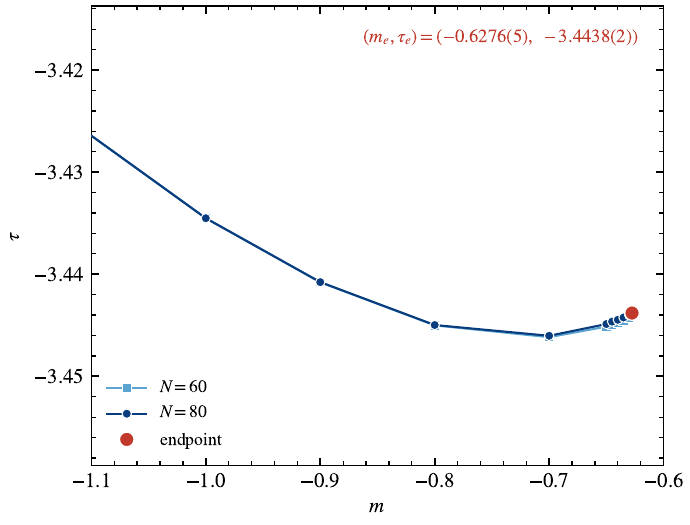}
    \caption{The line of first order phase transitions and its Ising fixed point at $x=100$. For this computation, the bond dimension was $\chi=500$, $\ell_{\rm max}=3$, and the symmetry breaking parameter $h=10^{-4}$.
}
    \label{fig:line_1st}
\end{figure}

Again we work at fixed $x=100$. We find that the line of first order phase transitions ends at the point $m_c= -0.6276(5)$ and $\tau=-3.4438(2)$, see Figure~\ref{fig:line_1st}. Using the same methods discussed in the previous section, we again find the endpoint to be of the Ising universality class, with a measured central charge of $c=0.497(10)$.

\subsection{The tricritical point}
We now focus on the determination of the tricritical point. We find that the computation with PBC works better in this case, as the range of values of $N$ we can reach appears far from the asymptotic region in the OBC case. Let us fix $x=100$, remarking once again, that this (as well as neglecting higher powers of $B$ in the generalized YM potential) can only affect nonuniversal quantities, such as the position of the tricritical point, and not the universal critical exponents upon tuning.

\begin{figure*}   \includegraphics[width = \linewidth]{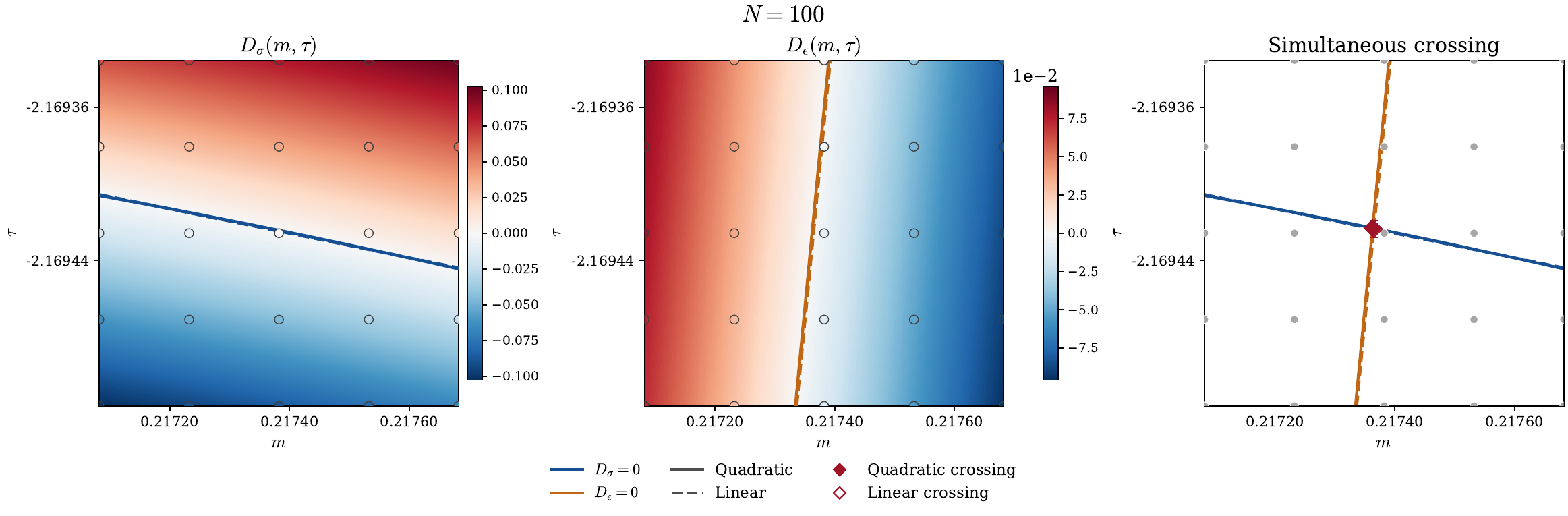}
    \caption{Determination of $\left(m_\times(100), \tau_\times(100)\right)$. We have interpolated the functions $D_\sigma,D_\epsilon$ in \eqref{eq:drifts} for $N=100$ on the $5\times 5$ rectangular grid of points $\circ$ . The intersection of the two curves $D_\sigma = D_\epsilon=0$ determines  the tricritical point $\left(m_\times(100), \tau_\times(100)\right)=(0.217363(7),-2.169423(4))$. We report the result of a quadratic and linear interpolation in this box, resulting in the same estimation. We have used $\chi =500$ and $\ell_{\rm max}=3$ for these determinations.}
\label{fig:diamondscrossing}
\end{figure*}

\begin{figure*}
    \centering
    \includegraphics[width =\linewidth]{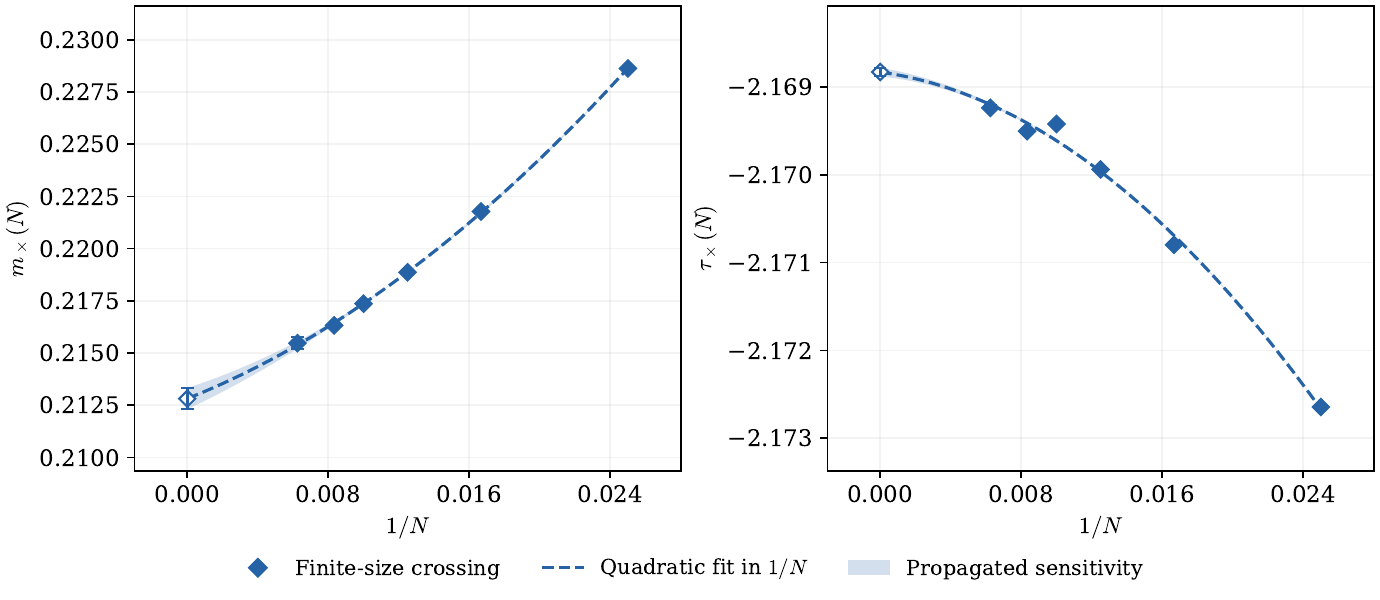}
    \caption{Extrapolation of the tricritical crossing positions using unweighted quadratic fits in $1/N$, where $N$ is the smaller size of the pair $(N,N+20)$. The fitted sizes are $N=40,60,80,100,120,160$. Open diamonds mark the $N\to\infty$ extrapolations. }
    \label{fig:mctc}
\end{figure*}

\begin{figure*}[ht!]
    \centering
\includegraphics[width=\linewidth]{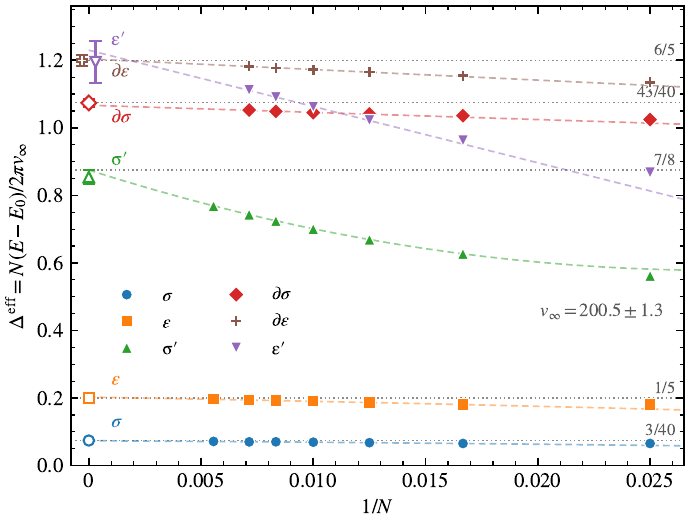}
    \caption{First low-lying states of the Hamiltonian with PBC. Here, $\ell_{\rm max} =3$, $\chi=500$ for $N <140$ and $\chi=1000$ for $N \geq 140$.  The extrapolation uncertainty is estimated from the envelope
of linear and quadratic spectrum and velocity fits with
$N_{\min}=40,60,80$, refitting the velocity for each window.
Extrapolated $N \to \infty$ values denote the midpoint of this envelope and
error bars its half-width.}
    \label{fig:tricriticalspectrum}
\end{figure*}

To identify finite-size candidates for the tricritical point, we use
the two lowest excitation gaps with PBC,\footnote{We used $\ell_{\rm max}=3$, $\chi =500$ for $N<140$, and $\chi =1000$ for $N\geq 140$.}
$\Delta E_\sigma=E_1-E_0$ and
$\Delta E_\varepsilon=E_2-E_0$, whose state assignments are
subsequently examined through the spectrum.
For each pair of sizes $(N,N+\delta N)$, with $\delta N =20$, we scan the bare couplings
$(m,\tau)$ and evaluate the size drifts for $i = \sigma, \epsilon$
\begin{equation}
\begin{split}
\label{eq:drifts}
D_i(m,\tau;N)
=&
(N+\delta N)\Delta E_i(N+\delta N;m,\tau)\\
&-
N\Delta E_i(N;m,\tau)
\end{split}
\end{equation}
We fit these measured drifts as smooth quadratic functions of both couplings by evaluating the energy levels with PBC on a two-dimensional discrete lattice of points $(m,\tau)$ ($5\times 5$ in Figure~\ref{fig:diamondscrossing}). This is illustrated in Figure~\ref{fig:diamondscrossing}.
 We determine the putative tricritical point at fixed $N$
\be
\left(m_\times(N), \tau_\times(N)\right)
\ee
by finding the intersection of the zero contour loci
$D_\sigma(m_\times(N),\tau_\times(N),N)=D_\varepsilon(m_\times(N),\tau_\times(N),N)=0$.
We have then repeated this for each consecutive pair among $N=40,60,\cdots,180$, giving $\left(m_\times(N), \tau_\times(N)\right)$ for each $N$.
Then, at each of these selected points we compute the spectrum of the model with PBC for the larger $N$ of the pair. The results are presented in Figure~\ref{fig:tricriticalspectrum}, reporting the first few energy levels per value of $N$. From the results above we can also extrapolate the tricritical point location as $(m^*_\times,\tau^*_\times)\equiv \lim_{N\to\infty} \left(m_\times(N), \tau_\times(N)\right) $
\be \begin{split}
(m^*_\times,\tau^*_\times)=\left(0.2128(5),-2.16883(5)\right)\end{split}\ee
 as in Figure~\ref{fig:mctc}.

 Figure~\ref{fig:tricriticalspectrum} reports the finite size spectrum at each of the points $N$.   We find very good agreement between the extrapolated $N \to \infty$ spectrum and what is expected for the tricritical Ising CFT
\begin{center}
\setlength{\tabcolsep}{12pt}
\begin{tabular}{ccc}
    \toprule
    State & Extrapolation & TCI \\
    \midrule
    $\sigma$         & $0.075\pm0.003$ & $0.075$ \\
    $\epsilon$         & $0.206\pm0.006$ & $0.200$ \\
    $\sigma'$          & $0.85\pm0.03$   & $0.875$ \\
    $\partial\sigma$   & $1.07\pm0.02$   & $1.075$ \\
    $\epsilon'$        & $1.24\pm0.03$   & $1.200$ \\
    $\partial\epsilon$ & $1.20\pm0.02$   & $1.200$ \\
    \bottomrule
\end{tabular}
\end{center}
showing very good agreement with the expected values. 
Although $\varepsilon'$ and the first descendants of $\varepsilon$
share the scaling dimension $\Delta=6/5$, $\varepsilon'$ is a scalar,
whereas $\partial\varepsilon$ and $\bar\partial\varepsilon$ (and likewise all the other first descendants) form a
left- and right-moving pair with conformal spins $s=\pm1$, corresponding
to momenta $p=\pm2\pi/N$ on the cylinder. We have reported only one of the two states with opposite momenta since they have identical energy in PBC.  This allows us to distinguish the two states.


\begin{figure}[t]
    \centering        \includegraphics[width=\linewidth]
{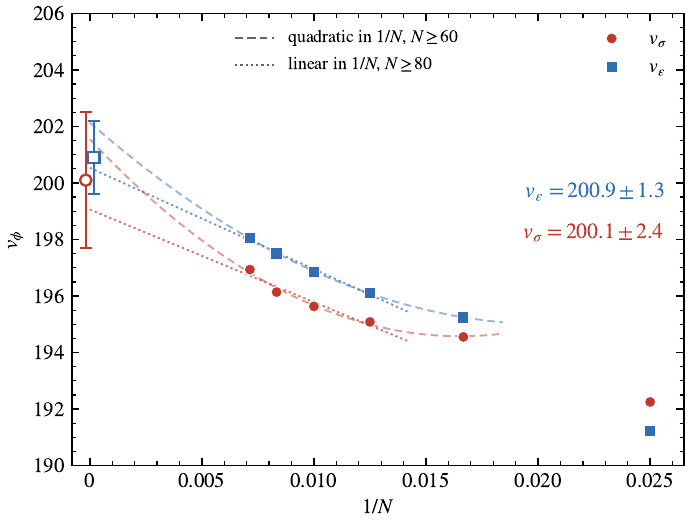}
    \caption{
    Extrapolation of the speed of light in the model using the spacing between the primary $\sigma$ and its descendant $\partial \sigma$ (red) and $\eps$ and $\partial \eps$ (blue).
    }
    \label{fig:pbc-tci-velocities}
\end{figure}

Using the states reported in Figure~\ref{fig:tricriticalspectrum}, we can follow the same procedure described above to determine the central charge as well. First we extract
the Casimir coefficient from the ground-state energy density,
fitting as in eq.\ \eqref{eq:cv}
resulting in\footnote{For this we need to fit at fixed value of $(m_\times,\tau_\times)$. We have reported here the value for  $(0.220, -2.1695 )$, that is consistent within the error with all the values extracted from all the points at fixed $(m_\times(N),\tau_\times(N))$ varying $N$.  }
\be 
cv=143.3 \pm 1.5
\ee
We can determine  
the velocity separately from the $\partial\sigma,\sigma$ and
$\partial\varepsilon,\varepsilon$ pairs as before. This gives (see Figure~\ref{fig:pbc-tci-velocities})
\be
v_\sigma=200.1 \pm 2.4, \qquad v_\eps = 200.9 \pm 1.3
\ee 
so we choose $v=200.5\pm 1.3$.  Note that this value of the velocity $v=2xu$ is compatible with $u=1$ as expected for a Lorentz invariant theory. 
This results in \be c=0.715 \pm 0.009\ee
consistent with the tricritical Ising value
$c=7/10$.

\section{Conclusions}
In this work, we have studied the generalized Schwinger model with a quartic potential for the $B$ field. 

The method used for the present work is to integrate out the gauge field and study a Hamiltonian that, in the case of OBC, only depends on fermions, but includes long-range interactions. This allows us to work directly on the gauge-invariant Hilbert space, which is finite dimensional at finite $N$. For PBC, one is left with an extra link variable which can take infinitely many values, which require truncation. In any case, these Hamiltonians can be represented with reasonable bond dimensions. From this point of view, it would be interesting to use the techniques of \cite{Dempsey:2025wia}, which generalize tensor network techniques in order to allow them to act on virtual indices as well rather than on physical indices only, avoiding the non-locality problem which would increase the MPO bond dimension; this would be useful in the case where more complicated potentials such as $B^6$ are considered.

We conjectured the phase diagram of the model first using semiclassical methods and then checked it numerically by discretizing the system. What we found was a region of unbroken charge conjugation separated from a region where this symmetry is spontaneously broken by a line of phase transitions. This line is composed of a line of first order and a line of second order phase transitions; these two lines meet at a tricritical fixed point, with central charge $c=7/10$. We also observe infinitely many lines of first order phase transitions inside of the C-broken region, which end at Ising CFTs.  When doing the numerical computations, we have worked at finite lattice spacing. As we approach the continuum limit, we expect the features to move in the phase diagram, but we don't expect the nature of the critical points, and therefore their critical exponents, to change. Therefore the computation of the finite lattice spacing phase diagram provides a check that the conjectured phase diagram is qualitatively correct.

Here, we limited ourselves to considering the very special case of a quartic B-field potential. Nothing prevents us from considering other choices. We expect the situation to be analogous: the phase diagram would consist of several critical,  multicritical as well as first order phase transitions. Let us comment on the case of odd potentials. In this case we should expect that an interesting role is played by $\mathcal{PT}$-symmetry at special values of the coupling. E.g.\ for the potential $V(B)= \tau B^2 + i B^3$, that is $\mathcal{PT}$-symmetric we expect to encounter the same phenomenology as in the Lee-Yang edge singularity \cite{PhysRevLett.40.1610,PhysRevLett.54.1354,Xu:2022mmw}. Finally, attempts at continuing two dimensional theories to non-integer dimensions have encountered serious obstacles \cite{DeCesare:2025ukl,DeCesare:2026dwm,Zan:2026oyb}. A proposed continuation of $SU(N)_1$ WZW models using the $N$-flavor Schwinger model was brought forward to \cite{Zhou:2022pah}. If such an approach would prove succesful, the generalization of the Schwinger model we discussed here could help in obtaining a continuation of multicritical fixed points to higher dimensions.

\begin{acknowledgments}
We are grateful to the organizers and participants of the workshops `XXI Avogadro meeting' at Universit\`a di Catania and the `YITP-IAS workshop: Interfaces \& Symmetry' (YITP-I-25-04) at Yukawa Institute for Theoretical Physics at Kyoto University, where part of this work was completed.
Research at Perimeter Institute is supported in part by the Government of Canada through the Department of Innovation, Science and Economic Development Canada and by the Province of Ontario through the Ministry of Colleges and Universities.
The research of BZ is supported by the Italian Ministry of University and Research under the Fondo Italiano per la Scienza (FIS3) programme, project QFTLand (CUP D53C25002660001).

\end{acknowledgments}
\bibliographystyle{apsrev4-2}
\bibliography{references}

\end{document}